\documentclass[12pt,preprint]{aastex}
\begin {document}
\title{{\bf Eccentricity Evolution in Simulated Galaxy Clusters}}
\author{Stephen N. Floor\altaffilmark{1,4}, Adrian L. Melott\altaffilmark{1},
Christopher J. Miller\altaffilmark{2}, and
Greg L. Bryan\altaffilmark{3}}
\altaffiltext{1}{Department of Physics \& Astronomy, University of Kansas, Lawrence, KS 66045}
\altaffiltext{2}{Department of Physics \& Astronomy, Carnegie Mellon University, Pittsburgh, PA 15213}
\altaffiltext{3}{Department of Astrophysics, Oxford University, Oxford, UK OX1 3RH}
\altaffiltext{4}{snfloor@ku.edu}

\newpage
\begin{abstract}
Strong cluster eccentricity evolution for $z \le 0.13$ has appeared in
a variety of observational data sets. 
We examine the evolution of eccentricity in simulated galaxy clusters using
a variety of simulation methodologies, amplitude normalizations, and
background cosmologies.  We do not find find such evolution
for $z < 0.1$ in any of our simulation ensembles.
We suggest a systematic error in the form of a redshift-dependent
selection effect in cluster catalogs or missing physics in cluster
simulations important enough to modify the cluster morphology.
\end{abstract}
\newpage

\section{Introduction}

Galaxy clusters are the largest bound objects in the Universe.  They
provide information on gravitational instability, and understanding their
formation is helpful in describing the transition to nonlinearity
in structure formation.

In studies of a variety of optical and X-ray samples, Melott, Chambers, and
Miller (2001, hereafter MCM) and Plionis (2002) found significant evolution in the gross 
morphology of galaxy clusters over the redshift range 0 to 0.13.  
When projected galaxy clusters were fit to ellipses by a variety of procedures,
their eccentricities were found to evolve on a timescale comparable to
one or two crossing times for the clusters.
The inclusion of X-ray data is important, as it is much less subject to
projection effects than optical data.
In all cases, with varying significance, the clusters were found to have
become less eccentric at lower $z$.  It was
argued that this could be understood as a relaxation process in a low-density
cosmological background, made possible by a greatly reduced recent merger
rate.
Mergers tend to occur by infall along supercluster filaments (Shandarin
and Klypin 1984; Colberg et. al 1999) which can lead to
alignment of the cluster axis (Bingelli 1982; Chambers, Melott, \&
Miller 2002, and references therein)
as well as aligned bulk flows within the cluster
(Novikov et al. 1999).

This poses new questions of structure formation for cosmological simulations.  Will cluster eccentricity evolution
appear there?  Does it depend on the initial conditions, the background
cosmology, and/or the physics included within given simulations?

Richstone, Loeb, \& Turner (1992) showed that cluster morphology could
be used as a diagnostic of mass density in the Universe.  While a 
cosmological constant, $\lambda$, was argued to have little effect,
a large mass density $\Omega_m$ would produce more irregular clusters due
to recent formation; on this basis they argued $\Omega_m \ge 0.5$.
Mohr et al. (1995) looked at a variety of simulations, finding their
$\Omega_m = 1$ simulations closest to the distribution of axial ratios
for observed clusters, since low-$\Omega_m$ simulations produced too-spherical
clusters.  Jing et al. (1995) disagreed, noting that open or $\lambda$-dominated
flat cosmologies also produced substantial irregularity in clusters.
Using the terminology OCDM and $\lambda$CDM for these, and $\tau$CDM for
the $\Omega_m = 1$ case (with the same initial conditions), they found increasing irregularity and eccentricity
in the order OCDM, $\lambda$CDM, $\tau$CDM.
They found the effect of $\lambda$ was smaller for
increasing $\Omega_m$, as one would expect.  They did not explore evolution.
Thomas et al. (1998) reinforced these conclusions.  Suwa et al. (2003)
presented results on ellipticity evolution, comparing clusters only at
$z=0$ with those at $z=0.5$ (but not intermediate redshifts).
These results suggest that OCDM clusters are slightly rounder than
$\lambda$CDM, although again the difference is small and not significant
in their study.
They found a small amount of ellipticity evolution between the two
redshifts.

We are interested in exploring the evolution of eccentricity in the recent past, 
especially for redshifts $0 \le z \le 0.1$, where there is a variety 
evidence for evolution in the data.
For this reason we have emphasized the outer regions of clusters, and used
methods similar to those applied to many observed clusters.

\section{Simulations}

We examined projected cluster properties in a variety of simulations
based on Cold Dark Matter and Gaussian perturbations (hereafter CDM).
One set has been previously described and extensively used elsewhere
(Loken et al. 2002, and references therein).
We refer to this as the $\lambda$CDMH simulation, to distinguish its inclusion
of hydrodynamics from other, purely N-body simulations we studied.
This simulation focused on the most massive objects within a region 256
$h^{-1} Mpc$ on a side in a flat, cosmological-constant dominated
universe ($\Omega_0 = 0.3$, $\Omega_\lambda= 0.7$) with a baryon
fraction of $\Omega_b=0.026$ and a Hubble constant of 70 km/s/Mpc.
The initial conditions used the Eisenstein \& Hu (1999)
parameterization of the CDM power spectrum, normalized
to an amplitude corresponding to $\sigma_8$=0.93, where $\sigma_8$
is the rms density fluctuation inside spheres of radius 8 $h^{-1} Mpc$.
The dark matter particle mass was $1.3 \times 10^{10} 
M_\odot$ and the evolution of the gas was followed with an Adaptive
Mesh Refinement method (see Bryan 1999) that solves the equations of
hydrodynamics (not including radiative cooling, thermal conduction, or star formation) on a
grid.  The method starts off with a relatively coarse mesh but
decreases the mesh spacing in dense region so that the smallest cell
size (best resolution) achieved was 16 $h^{-1}$ kpc.  We used X-ray
images assuming a metallicity of 0.3 of solar in the 0.5-2.0 keV
bandpass.
We used previously saved data on 11 clusters
taken at $z$ = 0.25, 0.1, and 0,
identified using the ``HOP" algorithm (Eisenstein and Hut 1998, hereafter EH).
This algorithm finds regions above some threshold density, and merges
them, associating them with regions of some maximal density $\delta_{peak}$.
These regions are merged into a single region (``cluster", for our purposes)
if they lie within a contour of density $\delta_{saddle}$.  Any associated
particles are also considered part of the cluster if they lie within a connected
contour above the density $\delta_{outer}$.
The parameters used to identify these clusters were
were 480, 400, and 160 for $\delta_{peak}$,
$\delta_{saddle}$, and $\delta_{outer}$, respectively.  For more details
on this method see EH.
In the $\lambda$CDMH, the mesh-refinement method requires following the same
volume element, chosen by the presence of a cluster at $z=0$, through
its evolution.  Clusters at higher redshift are taken from this volume; it
is possible that in this case they may not be the most massive clusters in
the large simulation volume at earlier times.
Another substantial difference with the larger cluster sample which will be analyzed next is that these very massive clusters
correspond to a Richness in excess of 2.  In fact they are even more massive than typical $R=2$ clusters, but less than $R=3$
clusters.  
It is not possible to change this given the AMR method, although
checking the conclusions against
other (N-body) simulations constitutes a partial cross-check.

We conducted another, larger, ensemble of pure Nbody simulations.
All used the same initial power spectrum, corresponding to CDM
(Bardeen et al. 1986)
with $\Gamma$=$\Omega$h=0.225.  This was evolved in both an open
($\Omega_{m0}$=0.34, OCDM1), a flat matter-dominated ($\Omega_{m0}$=1, $\tau$CDM),
and a flat cosmological-constant model
($\Omega_{m0}$=0.34, $\Omega_{\lambda 0}$=0.66, $\lambda$CDM1)
Inclusion of the $\tau$CDM
model was motivated by the wish to see whether major differences in the
expansion rate affected the eccentricity.   All three of these ensembles
were run as before, to an amplitude corresponding to $\sigma_8$=0.93.
A Hubble Constant $h=2/3$ was used in the
simulation analysis.  These N-body runs had three realizations each of $256^3$ particles in 
boxes of 128 $Mpc$ diameter. 
Given the volume and target space density of $R \ge 1$ clusters,
6 clumps were extracted from each of these three simulation "boxes".
For consistency with
our previous observational analyses, we simulated clusters of Richness 
$R \ge 1$.
Volume-limited samples exist for such clusters, but not for Richness 0.
We used parameters 400, 200, 80 for HOP in this case, in analogy with 
densities of observed clusters and virialization density thresholds.
Having identified clumps with HOP, clusters were identified with the
$N$ most massive clumps in the
simulation corresponding to a mean separation of $50 h^{-1}$ Mpc.  We
therefore have a total of 18 clusters from our three realizations.
Given two independent axial ratios out of three projections, we have
54 views at each redshift which have the statistical weight of 36
views.
In other words, this simulation ensemble produces the equivalent of 36
independent observed clusters for each assumed background cosmology.
(We have the equivalent weight of 22 views in the $\lambda$CDMH runs.)
The dark matter particle mass was $5.2 \times 10^{9} M_\odot$ in the low
$\Omega_m$ model and proportionally larger in the high $\Omega_m$ model.
Data was taken at moments
corresponding to $z$= 0.2, 0.1 and 0 in the OCDM1 and $\tau$CDM simulations.

In the OCDM1 boxes, our selected objects
ranged from about 8 $\times$ $10^{13}$ to 1 $\times$ $10^{15}$ M$_\odot$,
for the mass within an Abell radius.
This agreement with the mass range of observed clusters is a consistency
check on our procedure.
In the $\tau$CDM box, they were about three times more massive, as expected for
consistency with the assumed background cosmology.
In the $\lambda$CDMH box, the range was 
from about 5 $\times$ $10^{14}$ to 2 $\times$ $10^{15}$ M$_\odot$,
appropriate for $R > 2$ clusters.

Having chosen these clusters, three Cartesian projections were constructed.
There are only two independent axial ratios, $a/b$, and $b/c$, therefore only
two independent ratios in projection, whether we project along cluster principal 
axes or not.  We project along the simulation coordinate axes, since we do
not want to prefer principal axes; we use all 3 axes for balance.  In order
to take account of this partial lack of independence, we recognize that the
standard deviation of our results is smaller by a factor of
$\sqrt[]{\frac{2}{3}}$
than if they were fully independent.  This is taken into account in all
statistical results quoted here.

After initial analysis of OCDM1, we decided to examine more closely the
possible effect of amplitude normalization.  We therefore reran the 
entire ensemble with lower values of $\sigma_8$=0.78 and 0.58,
called OCDM2 and OCDM3 respectively.
Naturally, these clusters are somewhat less massive than in OCDM1.
Likewise, $\lambda$CDM2 has $\sigma_8$=0.83.

We have 33 projections of $\lambda$CDMH simulations at each redshift, and
54 of each of the OCDM*, $\lambda$CDM*, and $\tau$CDM simulations.
Including the fact that each of these is viewed at three redshift stages,
our simulated cluster sample size is larger than any of the individual data samples studied in
MCM, though not as large as the APM sample used by Plionis (2002).

Fair sampling is a consideration here.  We are viewing the same overall simulation
volume, and in most cases, the same clusters, at a variety of redshifts.
This is obviously not true in the observed samples. 
If this were important, cosmological simulations as currently analyzed would be of limited
relevance to understanding structure formation.  Separate simulations would
be needed for each redshift bin.  We partially compensate for
this effect (which we assume to be small)
in the OCDM*, $\lambda$CDM*,  $\tau$CDM simulations
by identifying clusters separately in each redshift bin.  This occasionally
resulted in the selection of different objects at different stages of evolution.

In the $\lambda$CDMH, the mesh-refinement method requires following the same
volume element, chosen by the presence of a cluster at $z=0$, through
its evolution.  Only the most massive clusters in this simulation were
chosen for mesh refinement.  The space density and threshold mass for the
objects chosen corresponds to a richness $R > 2$.
Clusters at higher redshift are taken from this sub-volume; it 
is possible that in this case they may not be the most massive clusters
at that $z$ in
the large simulation volume.
An additional caveat in the application of the $\lambda$CDMH group
is therefore its correspondence to richer clusters.
However, according to the study of Plionis (2002), ellipticity evolution
is not weaker for richer clusters.

\section{Determination of Eccentricity}

Observers had widely varying procedures for selecting which portions of 
clusters to analyze, and in some cases the procedure was not specified.
When major and minor projected axes were found, varying definitions of $\epsilon$,
the ellipticity, were applied.
In fact, varying definitions of ellipticity appear in common use in
mathematics, physics and astronomy literature.
We followed one well-specified procedure and
a standard definition of {\it eccentricity},
fitting ellipses to our projected clusters.
This geometrical term is well-known, and not subject to the ambiguity
in definitions of ``ellipticity".
For an ellipse with major and minor axes $a$ and $b$ respectively,
the eccentricity $e$ is defined by
\begin{eqnarray}
e = \sqrt[]{1 - \frac{b^2}{a^2}}
\end{eqnarray}
this definition of $e$ is also used for $\epsilon$ by some, but not all, work which reports 
on ``ellipticity".  Another common definition of ellipticity,
$\epsilon$= 1-b/a, has $e$ and $\epsilon$ increasing monotonically with one
another inside their useful ranges from $0$ to $1$.
Our statistical inferences, discussed later, are nonparametric and based on ranks; thus they would
be the same whether we used $e$ or $\epsilon$.

Only projected information is available for observed clusters in useful precision,
so we study our simulated clusters in projection.
The center of these objects was defined for most analyses here as 
the center of mass of the linked particles.  It must be stated clearly that
the procedures used in observational studies are not uniform, or even
always clearly specified.  This center of mass is our closest possible analog
for the center of mass of the optically luminous matter--the galaxies.
In the case of the $\lambda$CDMH simulation set, we also analyzed the 
synthetic X-ray emission of the clusters.  In this case, we chose as a
center the region of strongest simulated surface brightness in the X-ray.
For the sake of fuller exploration, we repeated this procedure using the
highest peak of the projected mass density as the center, and found no
qualitatively different conclusions after referring our
computation to the center of mass of regions drawn from around this center.

We selected the widely used ``annulus method";
it emphasizes the outer regions of the cluster, which is important
in trying to measure ellipticity for highly centrally concentrated X-ray images.
This is the most widely used method in cluster ellipticity work.
For each cluster projection, having chosen a center as described above,
a circle of radius 1.5 $h^{-1} Mpc$ was drawn
about this point.  Then another circle of smaller radius was placed so that
the annulus between them had 0.2 times the entire mass enclosed within the
outer circle.  The purpose of using annuli is to emphasize outer regions,
especially important for highly concentrated, emission-weighted X-ray images.
We repeated this procedure using 2.0 $h^{-1} Mpc$, with no qualitatively
different conclusions.
We also examined the eccentricity of our clusters when the entire mass within
1.5 $h^{-1} Mpc$ was used rather than an annulus.  Although eccentricity
values were generally lower, our conclusions again were qualitatively the
same.
We further tested the robustness of our conclusions against procedure by dropping the
annulus procedure, and computing the values based on the entire region connected by the HOP
procedure.  This also shifted the values somewhat, but did not change our conclusions about
evolution.  The results quoted herein are based on the 
1.5 $h^{-1} Mpc$ annulus.

We computed the ``inertia tensor" $\bf I$ of this annulus with respect to its
center of mass (center of emission for X-ray images).
Very few observational studies mentioned that they used
the center of mass frame; we hope that they did since the results are nearly
meaningless without.
If not computed with respect to the center of mass, the eigenvalues have
no particular relationship to the axes of the object.
We used the conventional definition 
\begin{eqnarray}
I_{ij} = \Sigma m_{\alpha} (\delta_{ij}r^2 - x_{\alpha i}  x_{\alpha j})
\end{eqnarray}
where we sum over regions $\alpha$.

The eigenvalues of $\bf I$ are proportional to $a^2$ and $b^2$, as defined above,
for a homogeneous ellipse.  In this way we define the eccentricity of
the projected simulated cluster approximated as an ellipse.

\section{Results}

In Table 1 below, we show for for the $\lambda$CDMH simulation,
the median and mean $e$ at each redshift bin, as well as the
standard deviation of the mean.
Note that the standard deviation in the estimate of the mean is 
smaller than the standard deviation of one measurement.
Figures 1 and 2 contain histograms of the eccentricity for the
OCDM1, $\tau$CDM, and $\lambda$CDMH ensembles.
Note that in $\lambda$CDMH, we show eccentricity inferred both
from the dark matter distribution, which is a stand-in for the optical
studies, assuming that galaxies reasonably trace the dark matter;
we also show results derived from the inferred X-ray luminosity.
There are weak evolutionary trends, but none as steep as that seen
in the data.
Again, note that this data sample corresponds to very rich clusters.

In agreement with previous studies, we find weakly increasing eccentricity in
the order OCDM1, $\lambda$CDM1, $\tau$CDM at $z=0$.

{\footnotesize
\begin{deluxetable}{ccccccc}
\tablewidth{0pt}
\tablecaption{$\lambda$CDMH Cluster Eccentricities, $N$=33, $R > 2$}
\tablehead{
\colhead{} & \colhead{} & \colhead{} & \colhead{} & \colhead{} & \colhead{} & \colhead{} \\
\cline{1-7} \\
\colhead{Simulation type} & \colhead{$\sigma_8$} &  \colhead{Redshift} &\colhead{Median $e$} & \colhead{Mean $e$} & \colhead{$\sigma_e$(mean)} & \colhead{$P_{W}$} }
\startdata
$\lambda$CDMH (mass) & 0.93   &  0.       &.63     & .61  &.027 & .24 (.16) \\
$\lambda$CDMH (mass) & --   &  0.1       &.66     & .63  &.028 & .46 \\
$\lambda$CDMH (mass) & --   &  0.25       &.68     & .64  &.026 & -- \\    
$\lambda$CDMH (X-ray) & 0.93   &  0.       &.61     & .66  &.036 & .33 (.34) \\
$\lambda$CDMH (X-ray) & --   &  0.1       &.73     & .68  & .033 & .65 \\
$\lambda$CDMH (X-ray) & --   &  0.25       &.70     & .67  & .029 & -- \\  
\enddata
\end{deluxetable}}

{\footnotesize
\begin{deluxetable}{ccccccc}
\tablewidth{0pt}
\tablecaption{N-body Cluster Eccentricities, $N$=54, $R \ge 1$}
\tablehead{
\colhead{} & \colhead{} & \colhead{} & \colhead{} & \colhead{} & \colhead{} & \colhead{} \\
\cline{1-7} \\
\colhead{Simulation type} & \colhead{$\sigma_8$} &  \colhead{Redshift} &\colhead{Median $e$} & \colhead{Mean $e$} & \colhead{$\sigma_e$(mean)} & \colhead{$P_{W}$} }
\startdata
OCDM1   & 0.93 &  0.       &.75     & .74  &.018 & .27 ($5 \times 10^{-5}$) \\
OCDM1   & -- &  0.1       &.77     & .76  &.019 & $4 \times 10^{-4}$ \\
OCDM1   & -- &  0.2       &.89     & .85  &.016 & -- \\  
$\lambda$CDM1   & 0.93 &  0.    &.77     & .78  &.014 & .045 ($7 \times 10^{-4}$)\\
$\lambda$CDM1   & -- &  0.1       &.83     & .80  &.020 & $ .033 $\\
$\lambda$CDM1   & -- &  0.2       &.88     & .85  &.015 & -- \\
$\tau$CDM& 0.93   &  0.       &.80     & .79  &.014 & .12 (.036) \\
$\tau$CDM& --  &  0.1       &.86     & .81  &.018 & .19 \\
$\tau$CDM& --  &  0.2       &.87     & .83  &.016 & -- \\   
OCDM2& 0.78  &  0.       &.87     & .83  &.017 & .21 (.076) \\ 
OCDM2& -- &  0.1       &.88     & .85  &.015 & .57 \\ 
OCDM2& -- &  0.2       &.89     & .86  &.015 & -- \\ 
OCDM3& 0.58  &  0.       &.89     & .84  &.018 & .11 (.98) \\ 
OCDM3& -- &  0.1       &.90     & .85  &.017 & .99 \\ 
OCDM3& -- &  0.2       &.83     & .77  &.023 & -- \\ 
$\lambda$CDM2 & 0.83   &  0.    &.85     & .83  &.015 & .030 (.017)\\
$\lambda$CDM2& -- &  0.1       &.88     & .86  &.014 & $ .47 $\\
$\lambda$CDM2& -- &  0.2       &.89     & .86  &.016 & -- \\
\enddata
\end{deluxetable}}

We also performed a Wilcoxon rank-sum test (Lehman 1998) to compare
samples at different stages of evolution and with one another.
A nice feature of this test is that it will give the same results for comparing
measures which are nonlinearly related, but monotonically increasing functions
of one another, like $e$ and $\epsilon$.
In this
case, the null hypothesis is that the lower-$z$ sample does not have systematically
smaller eccentricity than the high-$z$ sample.  In the last column, we
give the confidence in rejection of eccentricity evolution (reduction with
lower $z$).  In particular, the number in this column is the probability
that the change in eccentricity seen in the simulations could result from random draws of
a cluster population which did not have decreasing eccentricity.
The first number given in the last column is the confidence resulting from comparison with
the cluster ensemble on that line with the ensemble on the following line (next
higher redshift datafile).  The following number (in parentheses) is
the confidence resulting from comparison with the ensemble two redshift
bins higher (i.e. between redshift 0 and 0.2 or 0.25)
Small values of $P_W$ correspond to the direction of evolution seen in the data.

Although there is a significant signal in a variety of
both optical and X-ray data samples
that galaxy clusters have undergone a morphological
evolution, becoming significantly more circular since $z$ = 0.1, we have not found this
in our ensemble of simulations.  
There is such evolution only in the OCDM1 and $\lambda$CDM1 simulation sample for $0.1 \le z \le 0.2$.
OCDM3 has evolution in the opposite sense over the same redshift interval.
Evolution found in data by MCM was found only for $z \le 0.1$; that found
by Plionis was for $z \le 0.18$ and in fact appeared concentrated in
lower redshift ranges.  It was confirmed (Plionis, personal communication)
that no significant ellipticity evolution appears in that sample for $z \ge 0.13$.
Despite the differences in codes and lesser differences in cluster selection
procedures, none of the simulated samples showed strong eccentricity evolution
over the crucial period
$0 \le z \le 0.1$.
Evolution similar to that found in most of the data would correspond roughly to
eccentricity dropping from about 0.8 to 0.7 during that period, which would
stand out with high significance in our simulations.

There is a slightly different situation with the APM cluster sample.
The evolution rate is slower there; it would correspond to mean eccentricity
changing from about 0.76 to 0.70 during the redshift interval 0.1 to 0.
(extrapolating its slope).  This is a dangerous extrapolation, since this
sample does not contain information for $z < 0.038$, and its rate of evolution
appears to steepen for lower $z$.  Such a steepening would be consistent
with the more rapid rate of evolution found at lower redshifts in all but
one of the MCM samples.

It should be mentioned that the APM cluster sample was selected in a 
somewhat different way than the controlled Abell/ACO sample used 
elsewhere.  These clusters were initially found by selecting regions of
a high overdensity in a smaller region, about half the diameter typically
used in the Abell catalog (Dalton et al. 1997).  This procedure might
produce a selection effect accounting for the difference.

To summarize, the APM sample would appear to be consistent with the rate
of evolution in many of our samples at about the 2$\sigma$ level, if
extrapolated linearly to $z$=0.

\section{Discussion}

In some way, the simulations as we have analyzed them do not faithfully
model the observed ``local" Universe.  There are a variety of ways to account for
this; we will mention a few.

We initially used a normalization $\sigma_8$=0.93.  There is a hint in
the OCDM1 results that a lower normalization might produce the
strong evolution seen in the data in the redshift range needed.
We decided to test this, and
the OCDM2 and OCDM3 results provide evidence against this hypothesis
(see Table 1 and Figure 3).
We also wished to investigate cosmic variance beyond the simple variation of
random phases in the initial conditions.  Density fluctuations on the scale
of the N-body box should be no more than about 10\% $rms$.  As an aggressive
test of cosmic variance, we reran OCDM1 in a background with $\Omega_{m0}$=0.2.
This also would be an effective test of any possible effects due to our
being located in a very large underdense region, such as that suggested
by Busswell et al. (2003) and Frith et al. (2003).  We found no significant
change in the results of the ensemble.

The most obvious possibility is observational selection effects.  The sign
of the effect needed is that more eccentric clusters are included in
cluster surveys when they are far away than when they are close.

The data used in MCM and Plionis (2002) are based on optical cluster
catalogs.  In MCM, they used subsets of the Abell/ACO catalogs that
had published ellipticities (in either X-ray or optical). Plionis (2002)
used a subset of the Automated Plate Machine (APM) clusters as published
in Dalton et al. (1997), with ellipticities as given in Basilikos,
Plionis, and Maddox (2000). Plionis (2002) also looked at Abell/ACO clusters
in their APM sample and found the same ellipticity evolution as for the
entire sample.

In the optical and Xray samples examined in MCM, ellipticity
appears to increase with redshift for all samples. This increase
exists in all samples studied in both papers, but the way
in which it appears is different for different samples:
(1) a lack of nearby highly elliptical
clusters (compared to higher redshift);
(2) an increase
in the number of highly elliptical clusters at higher redshift (compared
to nearby); (3) a lack of highly circular clusters at high-$z$
(compared to low-$z$); and of course any combination of these.
In fact, there are observed samples consistent with all of these possibilities
(MCM, Plionis 2002).
It is worthwhile asking whether selection effects (both at
the low-$z$ and the high-$z$ end of the cluster samples) could play a
role in this correlation.

For instance, it is
possible that both the APM and the Abell/ACO catalogs are incomplete
at low redshifts and that those clusters missing from the sample have
significant ellipticity. Likewise, 
an incompleteness in circular clusters at higher redshifts
is also possible. 
This latter case seems less plausible, since such ``regularly'' shaped clusters
should be easy to identify. Ebeling et al. (1999) claimed that high levels of
substructure will lead to a decreased probability of cluster selection.
The selection function for regularly shaped (circular)
clusters should be stable over the range
of redshifts examined in MCM and Plionis et al.
In this case it is doubtful that circular clusters would be preferentially
missed at higher redshifts.

However, Adami et al. (2000) conducted
extensive testing on their X-ray cluster selection function (for the
Serendipitous High-Redshift Cluster (SHARC) survey (Romer et al. 2000)). They reported that
highly elliptical clusters were {\it more easily} found at high redshift.
Adami et al. note that for a fixed cosmology, an elliptical cluster will
have a higher central surface brightness than a circular cluster with the
same luminosity. They found that this effect was only significant at the high-$z$ limit of
the SHARC survey, where the selection function is very sensitive to the central
surface brightness of a cluster.
All redshifts simulated by Adami et al. were greater than the ones considered
here.
In spite of this objection, it is possible that elliptical clusters
could be more easily
detected at high-$z$ and that some highly circular clusters could be missed at high-$z$.
Since all of the samples examined in MCM and Plionis et al. were optically selected,
one would have to assume that an analogous effect could be present in optical samples.

In general, it remains unclear how ellipticity affects the
selection of clusters. The argument made by Ebeling et al. (1999) that clusters
with substructure are harder to detect, need not necessarily translate
into elliptical clusters being harder to detect. The use of extensive simulations
(as in Adami et al. 2000) is probably a requirement before any conclusions
can be drawn regarding a ellipticity evolution in any given dataset.
Unfortunately, the selection procedures used for the cluster samples in Plionis (2002)
and MCM are so complicated that their selection functions
are too difficult to model. Our best hope is to examine the next generation of
cluster catalogs (e.g. from the Sloan Digital Sky Survey) whose completeness
and selections should be well-measured.

It is possible that the method of defining included mass and computing
eccentricity could affect these results.  However, we also explored using
a {\it friends of friends} algorithm, as well as various HOP parameters,
and abandoning the annulus method in favor of inclusion of all mass in
the clusters.  None of these changed the evolution of eccentricity, though
some changed its amplitude (consistently over redshifts, so as to preserve
the lack of evolution).

Another possibility is that the simulations are missing crucial physics
which might induce this change.  Although violent relaxation may 
isotropize objects through the time-dependent gravitational potential,
we cannot exclude processes in the hot gas which is the dominant baryonic
component of the clusters.  Radiative cooling may cause the gas to 
contract, gravitationally entraining some of the dark matter.
Also, it has been shown (Narayan and Medvedev 2001) that thermal conduction
may be important, especially in the turbulence following major mergers.

\section{Acknowledgments}

We thank M. Plionis and G. Evrard for useful comments.
SNF and ALM gratefully acknowledge the support of the National Science
Foundation through grant AST-0070702, especially a supplement for Research
Experiences for Undergraduates.
Invaluable computing support came from the National Center for Supercomputing
Applications.

\vfill \eject

\begin{figure}[t]
\epsscale{.7}
\plotone{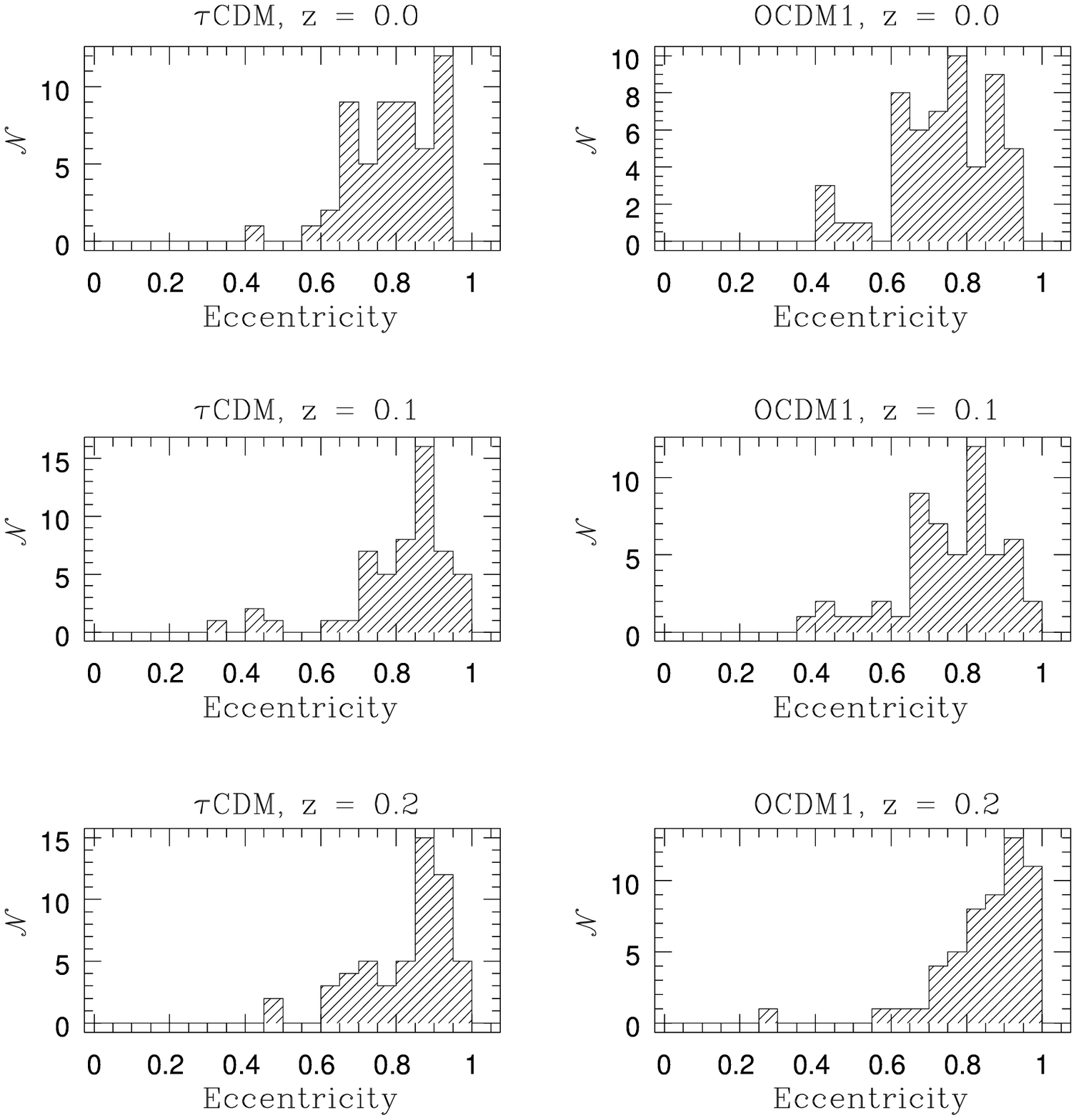}
\caption[]{\footnotesize
Distribution of eccentricity for the OCDM1 (low $\Omega$) and
$\tau$CDM (high $\Omega$) clusters at redshifts 0, 0.1, and 0.2.
}
\end{figure}

\begin{figure}[t]
\epsscale{.7}
\plotone{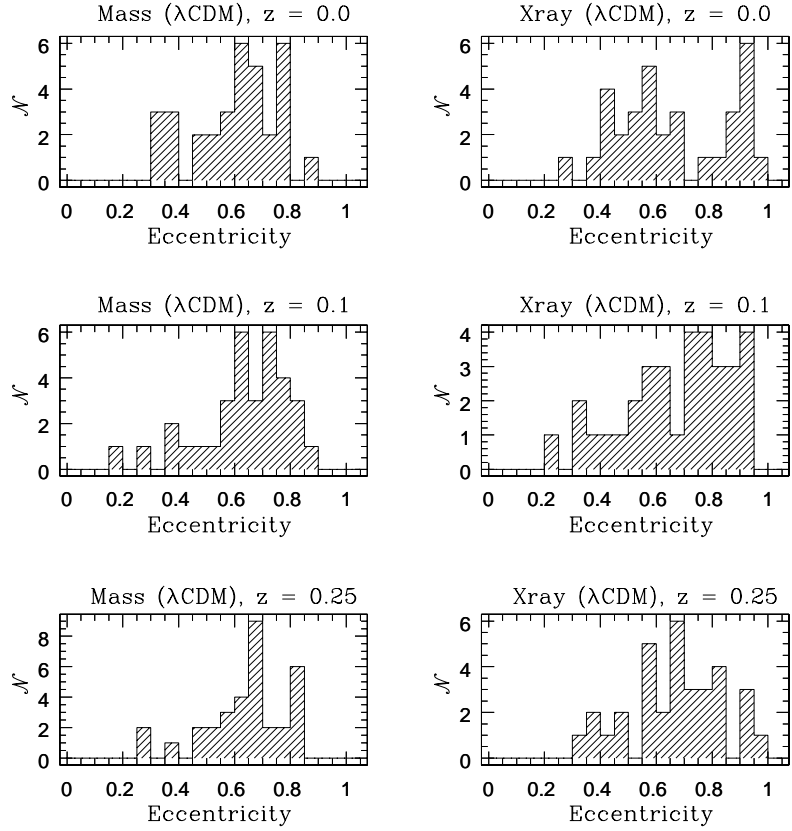}
\caption[]{\footnotesize
Distribution of eccentricity for the $\lambda$CDMH clusters at redshifts
0, 0.1, and 0.25.  The eccentricity of the mass distribution and of the
simulated X-ray emission are plotted separately.
}
\end{figure}

\begin{figure}[t]
\epsscale{.7}
\plotone{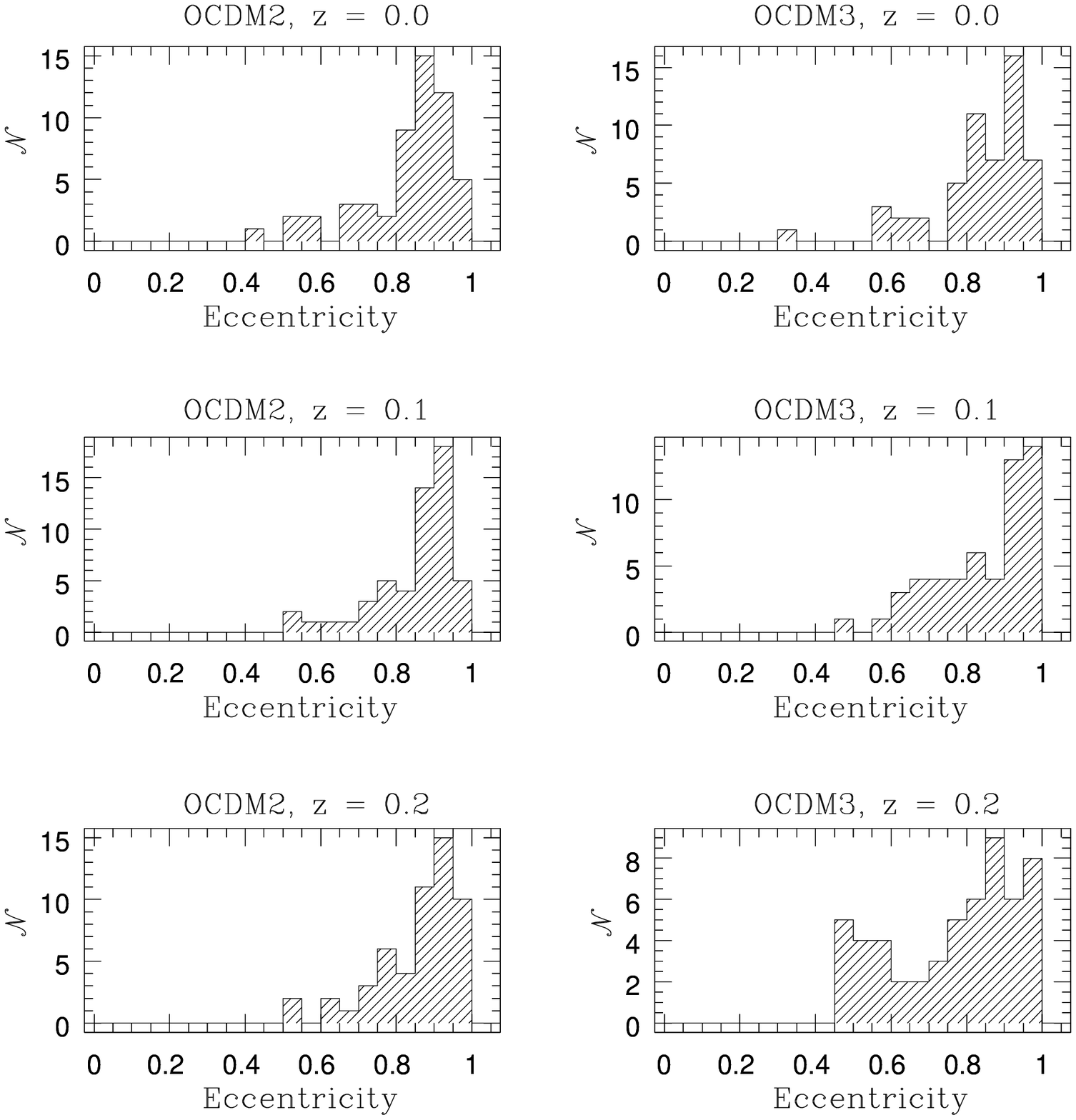}
\caption[]{\footnotesize
Distribution of eccentricity for the OCDM2 and OCDM3
(low $\Omega_m$, low normalization: see text) 
clusters at redshifts 0, 0.1, and 0.2.
}
\end{figure}


\begin{thebibliography}{}
\bibitem {2000ApJS..131..391A} Adami, C., Ulmer, M.~P.,
Romer, A.~K., Nichol, R.~C., Holden, B.~P., \& Pildis, R.~A.\ 2000, \apjs,
131, 391
\bibitem[BBKS]{bbks}Bardeen, J.M., Bond, J.R., Kaiser, N., \& Szalay,
A.S. 1986, \apj   304, 15
\bibitem[basi]{basi}Basilakos, S., Plionis, M., \& Maddox, S. 2000, \mnras, 316, 779
\bibitem[(Binggeli 1982)]{bing82}Binggeli, B. 1982, A\&A  107, 338
\bibitem[(bry)]{bry}
Bryan, G.L. 1999, Computing in Science and Engineering, 1:2, 46
\bibitem[buss]{buss}Busswell, G.S., Shanks, T., Outram, W.J., Firth, N.,
Metcalfe, N., \& Fong, R. 2003, astro-ph/0302330

\bibitem[CMM]{cmm}Chambers, W.C., Melott, A.L., \& Miller, C.J. 2002,
\apj, 565, 849
\bibitem[Colberg et al 1999]{col99}Colberg, J.M., White, S.D.M., Jenkins, A.,
\& Pearce, F.R. 1999, \mnras, 308, 593
\bibitem[Chambers et al 2002]{ch02}Chambers, S.W., Melott, A.L., \&
Miller, C.J. 2002, \apj, 565, 849
\bibitem[dal]{dal}Dalton, G. B., Maddox, S. J., Sutherland, W. J., \& Efstathiou, G. 1997, \mnras, 289, 263
\bibitem[Ebeling et al.(2000)]{2000ApJ...534..133E} Ebeling, H.~et al.\
2000, \apj, 534, 133
\bibitem[EH]{EH}Eisenstein, D.J., \& Hut, P. 1998, ApJ, 498, 137
\bibitem[Eis]{eis}Eisenstein, D.J., \& Hu, W. 1999, ApJ, 518, 2
\bibitem[Fir]{fir}Firth, W.J., Busswell, G.S., Fong, R., Metcalfe, N.,
\& Shanks, T. 2003, astro-ph/0302331
\bibitem[Jing]{ji}Jing, Y.P., Mo, H.J.,  B\"orner, G., \& Fang, L.Z.
1995, MNRAS, 276, 417
\bibitem[Lehman]{leh}Lehman, E.L. 1998, Nonparametrics (Upper Saddle
River: Prentice-Hall)
\bibitem[Lok]{lok}Loken, C. et al. 2002 ApJ 579, 571
\bibitem[Melott etal 2001]{mel01}Melott, A.L., Chambers, S.W., \&
Miller, C.J. 2001, \apj, 559, L75
\bibitem[Mohr]{mo}Mohr, J.J., Evrard, A.E., Fabricant, D.G.,
\& Geller, M.J. 1995, ApJ, 447, 8
\bibitem[NarMed]{narmed} Narayan, R., \& Medvedev, M.V. 2001
ApJ Letters, 562, L129
\bibitem[Novikov et al 1999]{nov99}Novikov, D.I., Melott, A.L,, Wilhite,
B.C., Kaufman, M., Burns, J.O., Miller, C.J., and Batuski, D.J. 1999,
\mnras, 304, L5.
\bibitem[Plionis 2002]{pli02} Plionis, M. 2002, \apj, 572, L67.
\bibitem[RLT]{ri}Richstone, D., Loeb, A., \& Turner, E.L. 1992 
ApJ 393, 477
\bibitem[(Shandarin \& Klypin 1984)]{shkl84} Shandarin, S.F., and Klypin, A.A.,
        1984, Sov Astron. 28, 491
\bibitem
[SHYO]{shyo}Suwa, T., Habe, A., Yoshikowa, K., \& Okamoto, T. 2003 ApJ, 588, in press
(astro-ph/0108308)

\bibitem[Thomas et al]{thom} Thomas, P.A., et al., 1998, MNRAS, 296, 1061

\end{thebibliography}
\end {document}